\documentclass[conference]{IEEEtran}

\usepackage{color}
\usepackage{comment}
\usepackage{cite}
\usepackage{multirow}
\usepackage{graphicx}
\usepackage{caption}
\usepackage{subcaption}
\usepackage{epstopdf}
\usepackage{verbatim}
\usepackage{amsmath}
\usepackage{hyphenat}
\usepackage{fancyvrb}
\usepackage{boxedminipage}
\usepackage{xcolor}
\usepackage{amsmath}
\usepackage{algorithm}
\usepackage[noend]{algpseudocode}
\usepackage{array}
\usepackage{amssymb}

\newcommand{\argmin}{\operatornamewithlimits{argmin}}

\makeatletter
\def\endthebibliography{%
	\def\@noitemerr{\@latex@warning{Empty `thebibliography' environment}}%
	\endlist
}
\makeatother

\begin{document}

\title{Accelerating MRI Reconstruction on TPUs}

\author{
	\IEEEauthorblockN{Tianjian~Lu\IEEEauthorrefmark{1}, Thibault~Marin\IEEEauthorrefmark{2}, Yue~Zhuo\IEEEauthorrefmark{2}, Yi-Fan~Chen\IEEEauthorrefmark{1}\IEEEauthorrefmark{3}, and  Chao Ma\IEEEauthorrefmark{2}\IEEEauthorrefmark{3}}
	\IEEEauthorblockA{\IEEEauthorrefmark{1}Google Research
		\\\{tianjianlu, yifanchen\}@google.com}
	\IEEEauthorblockA{\IEEEauthorrefmark{2}Gordon Center for Medical Imaging, Massachusetts General Hospital, Harvard Medical School
		\\\{tmarin, yzhuo2, cma5\}@mgh.harvard.edu}
	\IEEEauthorblockA{\IEEEauthorrefmark{3}Corresponding authors}
}

\markboth{}%
{Shell \MakeLowercase{\textit{Lu et al.}}: Accelerating MRI Reconstruction on TPUs}

\maketitle

\begin{abstract}
The advanced magnetic resonance (MR) image reconstructions such as the compressed sensing and subspace-based imaging are considered as large-scale, iterative, optimization problems.
Given the large number of reconstructions required by the practical clinical usage, the computation time of these advanced reconstruction methods is often unacceptable.
In this work, we propose using Google's Tensor Processing Units (TPUs) to accelerate the MR image reconstruction. 
TPU is an application-specific integrated circuit (ASIC) for machine learning applications, which has recently been used to solve large-scale scientific computing problems.
As proof-of-concept, we implement the alternating direction method of multipliers (ADMM) in TensorFlow to reconstruct images on TPUs.
The reconstruction is based on multi-channel, sparsely sampled, and radial-trajectory $k$-space data with sparsity constraints. 
The forward and inverse non-uniform Fourier transform operations are formulated in terms of matrix multiplications as in the discrete Fourier transform.
The sparsifying transform and its adjoint operations are formulated as convolutions.
The data decomposition is applied to the measured $k$-space data such that the aforementioned tensor operations are localized within individual TPU cores.
The data decomposition and the inter-core communication strategy are designed in accordance with the TPU interconnect network topology in order to minimize the communication time.
The accuracy and the high parallel efficiency of the proposed TPU-based image reconstruction method are demonstrated through numerical examples.
\end{abstract}


\section{Introduction}
Magnetic resonance imaging (MRI) is a powerful imaging tool that non-invasively reveals the structural, functional and biological information of the human body. Because of its excellent soft-tissue contrast and high spatial resolution, MRI has revolutionized the field of medical imaging since its invention in 1970s. Over decades of development in MR hardware and imaging sequences, the data acquisition speed of MR is approaching both the physical and physiological limits.
As a result, the Nyquist sampling criterion of the conventional Fourier-based image reconstruction becomes the bottleneck for a further acceleration of MRI.
Modern MR imaging methods such as parallel imaging \cite{sodickson1997simultaneous, pruessmann1999sense, pruessmann2006encoding}, compressed sensing \cite{candes2006robust, donoho2006compressed, candes2006near,lustig2007sparse}, and subspace-based imaging \cite{liang2007ps} significantly reduce the imaging time by sparsely sampling the $k$-space.
Artifact-free images are reconstructed from the undersampled data by leveraging additional spatial encoding offered by coil sensitivities and prior knowledge of the underlying MR signal (e.g., sparsity).
Using the advanced reconstruction methods to accelerate MR and alleviate the demands for novel MR hardware becomes a trend in this field.
However, the advanced reconstruction methods often build upon large-scale, iterative, optimization algorithms with extensive usage of non-uniform Fourier transform \cite{pruessmann2006encoding,lustig2007sparse}, the computation time of which is often unacceptable for practical clinical use. 

In order to achieve clinically practical runtime, hardware accelerators and parallel computing have been used to handle the computationally demanding MR reconstruction tasks \cite{stone2007gpumr,pratx2011gpumedicalphysics,eklund2013medicalgpu,despres2017reviewgpurecon,wang2018surveygpumr,Zhuo2011}.  Graphics processing units (GPUs) have been extensively studied to accelerate the reconstruction: first by accelerating the non-uniform Fourier transform, which is the main bottleneck in the reconstruction \cite{stone2008nufft,soerensen2008nufft,yang2009nufft,Zhuo2010d}. Further, iterative image reconstruction methods such as conjugate gradient (CG) and alternating direction method of multipliers (ADMM) \cite{boyd2011distributed} have been implemented on GPUs for compressed sensing problems \cite{hansen2008cartktsense,murphy2012l1spirit,smith2012bregman,nam2013cs3drad,chang2017cs3d} enabling advanced reconstruction from multi-channel undersampled data.  Several packages offer GPU-based iterative image reconstruction from non-Cartesian $k$-space data acquired with phased-array coils \cite{schaetz2012multigpulib,freiberger2013agile,wu2011impatient,gai2013moreimpatient,cerjanic2016powergrid,uecker2019}. 
Despite these significant advances, runtimes for large-size dynamic image reconstruction problems are still not compatible with clinical practice. The recent success of machine learning (ML), or deep learning (DL) in specific, has spurred a new wave of hardware accelerators \cite{stoica2017berkeley}, among which Google's Tensor Processing Unit (TPU) \cite{jouppi2017datacenter} is considered as a promising approach to address the computation challenge brought by the continuous and exponential growth of data. Even though TPU is designed as an application-specific integrated circuit (ASIC) to run cutting-edge ML models on Google Cloud \cite{tpuv3}, it has recently been employed to tackle large-scale scientific computing problems \cite{yang2019high, belletti2019tensor, lu2020large}.

In this work, we propose using TPUs to accelerate the MRI reconstruction. There are four major advantages of deploying MR image reconstruction on TPUs, which can be understood from the TPU system architecture and the reconstruction algorithm used in this work. The first advantage is owing to the fact that the major operations involved in MR image reconstruction, including the non-uniform Fourier transform, the sparsifying transform, and the encoding of sensitivity profiles can all be formulated as tensor operations. Formulating the reconstruction problem as tensor operations is to take full advantage of TPU's strength in the highly efficient matrix multiplications \cite{lu2020large,jouppi2017quantifying}.  The second advantage stems from the data decomposition and communication strategies, which coincide with the interconnect network topology of TPU. Therefore, all the aforementioned tensor operations are localized on individual cores and the minimal communication of an image-size data is required at each iteration. It is worth mentioning that the communication among TPU cores does not go through host CPUs or networking resources. The third advantage is a result of the large capacity of the in-package memory of TPU, which makes it possible to handle large-scale problems efficiently. As a reference, one TPU v3 unit (or board) contains four chips and provides 128 GiB high-bandwidth memory (HBM) \cite{tpuv3}. The forth advantage is associated with the fact that TPU is easily programmable with software front ends such as TensorFlow \cite{wu2016google}. The TensorFlow TPU programming stack can express the parallel computing algorithms with simple and easy-to-understand code. In addition, TensorFlow offers a rich set of functionalities for scientific computing, which smooths the path of deploying the distributed MR image reconstruction algorithms on TPUs. Owing to the aforementioned four advantages, the distributed MR image reconstruction on TPUs can achieve very high computation and parallel efficiency.

To demonstrate the accelerated MR image reconstruction on TPUs, we implement the ADMM algorithm for MR image reconstruction in TensorFlow.
More specifically, we apply the ADMM alorigthm to reconstruct images from multi-channel, sparsely sampled, and radial-trajectory $k$-space data with sparsity constraints. The forward and inverse non-uniform Fourier transforms are treated as matrix multiplications between the Vandermonde matrix and the image as in the discrete Fourier transform (DFT). The sparsifying transform and its adjoint operations are formulated as convolutions. Both the accuracy and the high parallel efficiency of the proposed TPU-based image reconstruction method are demonstrated with numerical examples.

\begin{figure}
	\centering
	\begin{subfigure}{\linewidth}
		\centering
		\includegraphics[width=6cm, height=2cm]{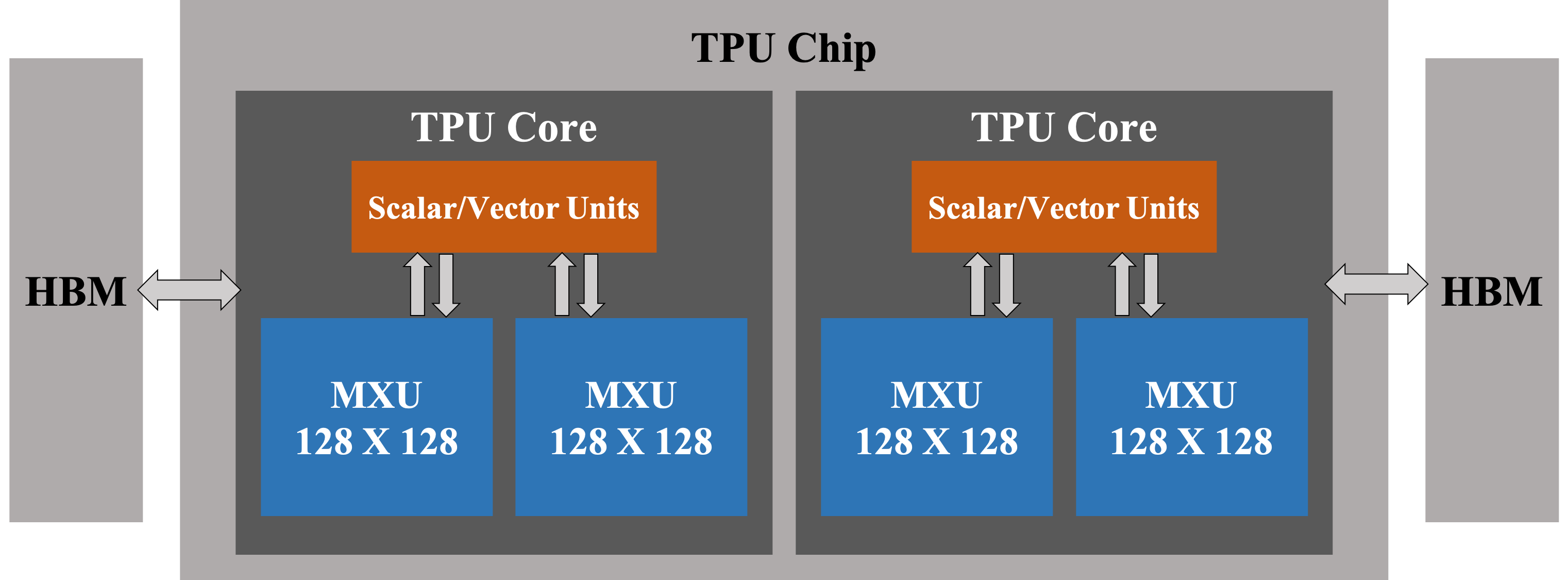}
		\caption{} 
	\end{subfigure}\vfill
	\begin{subfigure}{\linewidth}
		\centering
		\includegraphics[width=6cm, height=2.5cm]{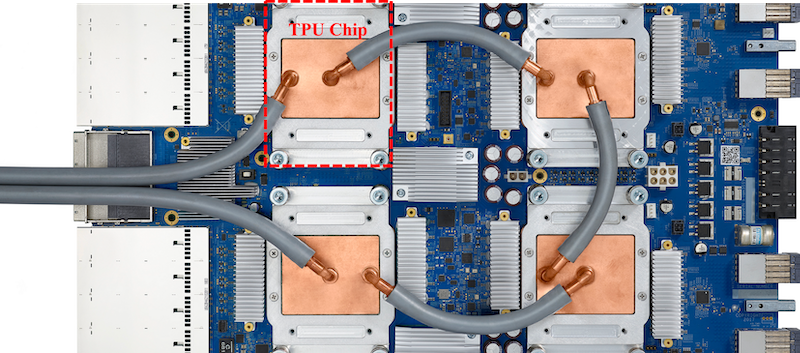}
		\caption{} 
	\end{subfigure}
	\begin{subfigure}{\linewidth}
		\centering
		\includegraphics[width=6cm, height=2.5cm]{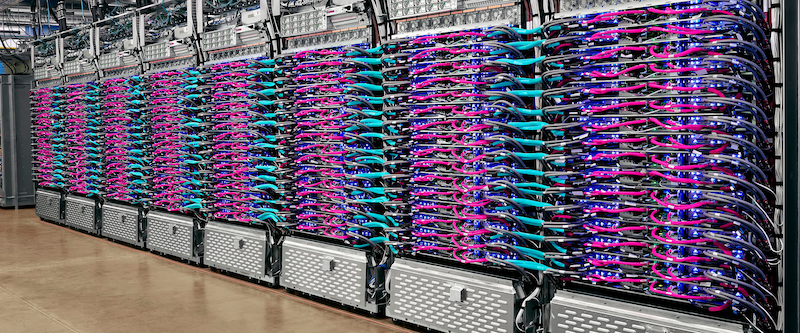}
		\caption{} 
	\end{subfigure}	
	\caption{The TPU v3 (a) chip, (b) unit (or board), and (c) Pod. One TPU board has four chips; each chip contains two cores; and a TPU v3 Pod in a data center contains 2048 cores.}
	\label{tpu_chip_board_pod}
\end{figure}

\section{TPU system architecture}
Understanding the advantages of deploying the MR image reconstruction  on TPUs cannot be separated from the knowledge of TPU system architecture. In this section, we provide an overview of the TPU system architecture on both the hardware and software components.

\subsection{Hardware architecture}
Figures \ref{tpu_chip_board_pod}(a) and (b) show one TPU chip and unit (or board), respectively.
One TPU unit contains four chips;  each chip has two cores; and each core contains the scalar, vector, and matrix units (MXU).
MXU provides the bulk of the compute power of a TPU chip and it handles 16 K multiply-accumulate (MAC) operations in one single clock cycle.
Both the inputs and outputs of MXU are float32, whereas MXU performs MAC operations with bfloat16 \cite{bfloat16}.
By leveraging the strategy of decomposition and accumulation, the image reconstruction in this work achieves the precision equivalent to float32.
As shown in Fig.~\ref{tpu_chip_board_pod}(b), each TPU core has 16 GiB HBM.
TPU is designed as coprocessor on the I/O bus: each board is paired with one host server
consisting of CPU, RAM, and hard disk; TPU executes the instructions sent from CPU on the host server through PCIe.

In a Pod configuration as shown in Fig.~\ref{tpu_chip_board_pod}(c), TPU chips are connected through dedicated high-speed interconnects of very low latency.
There are a total number of 2048 cores in a TPU v3 Pod.
The interconnect topology is a two-dimensional (2D) toroidal mesh with each chip connected to its four nearest neighbors.
As the interconnects are on the device, the communication among TPU cores does not go through a host CPU or any networking resource.

\subsection{Software architecture}
We program TPUs through TensorFlow.
Running a program on TPUs consists of four steps:
first, a TensorFlow client converts the TensorFlow operations into a computational graph and sends the  graph to a TensorFlow server;
second, the TensorFlow server partitions the computational graph into portions that run on TPU and CPU, respectively;
if multiple TPUs are to be employed, the graph is marked for replication;
third, the TensorFlow server  compiles the sub-graph that runs on TPUs into a high-level optimizer (HLO) program
and invokes the accelerated linear algebra compiler (XLA);
and last, the XLA compiler takes in the HLO program and converts it into a  low-level optimizer (LLO) program, which is effectively the assembly code for TPUs.
Both the generation and compilation of the computational graph occur on the host server.
The compiled LLO code is loaded onto TPUs for execution from the host server through PCIe.

\section{Image Reconstruction with ADMM}
In this section, we provide the details on the formulation of the image reconstruction problem in compressed sensing MR and the ADMM algorithm.

\subsection{Compressed sensing MRI}
A measured complex value $d_{\kappa, \gamma}$ at the $\kappa^{\mathrm{th}}$ position in the $k$-space by the $\gamma^{\mathrm{th}}$ coil can be expressed as
\begin{align}
d_{\kappa, \gamma} &= \sum_n  s_{n, \gamma} \rho_n e^{-i 2 \pi \mathbf{k}_{\kappa} \cdot \mathbf{r}_n} \\
&= \left[ \mathbf{F}(\boldsymbol{\rho}) \right]_{\kappa, \gamma},
\end{align}
where $\mathbf{r}_n$ denotes the spatial coordinate on a uniform grid, $\rho_n$ denotes the image intensity defined on $\mathbf{r}_n$, $\mathbf{k}_{\kappa}$ denotes the $k$-space coordinate on a non-uniform grid, $s_{n, \gamma}$ is the spatial sensitivity of the $\gamma^{\mathrm{th}}$ coil on $\mathbf{r}_n$, and $\mathbf{F}$ denotes the DFT operator.

In compressed sensing, we reconstruct an image from the undersampled $k$-space data by solving the following regularized least-square optimization problem \cite{candes2006robust, donoho2006compressed, candes2006near,lustig2007sparse}:
\begin{equation}
\underset{\boldsymbol{\rho}}{\mathrm{min}}\hspace{0.3cm}  \lVert \mathbf{F} \left(\boldsymbol{\rho}\right) - \mathbf{d} \rVert_2^2 + \lambda \lVert \boldsymbol{\Theta}(\boldsymbol{\rho}) \rVert_1
\label{bpdn_opt_unconstraint}
\end{equation}
where $\boldsymbol{\Theta}(\boldsymbol{\rho})$ denotes a sparsifying transform and $\lambda \!> \!0$ is the regularization parameter.
The first term of the cost function in (\ref{bpdn_opt_unconstraint}) is a data fidelity term and the second term is used to enforce the sparsity of the reconstructed image in a certain transformed domain. 
In this work, the finite-difference operator is used as the sparsifying transform.

\subsection{Reconstruction with ADMM}
The optimization problem in (\ref{bpdn_opt_unconstraint}) is a large-scale convex optimization problem, which can be solved efficiently using the ADMM algorithm \cite{ramani2010parallel}.
In ADMM, one introduces an auxiliary variable $\mu$ such that (\ref{bpdn_opt_unconstraint}) becomes
\begin{align}
&\underset{\boldsymbol{\rho}} {\mathrm{min}} \hspace{0.3cm} \lVert \mathbf{F} \left(\boldsymbol{\rho}\right) - \mathbf{d} \rVert_2^2 + \lambda \lVert \boldsymbol{\mu} \rVert_1 \nonumber  \\
&{\mathrm{s.t.}} \hspace{0.3cm} \boldsymbol{\Theta}\left( \boldsymbol{\rho} \right) - \boldsymbol{\mu} = 0.
\label{admm_obj}
\end{align}
The augmented Lagrangian associated with (\ref{admm_obj}) can be written as
\begin{align}
L_{\beta} \left(\boldsymbol{\rho}, \boldsymbol{\mu}, \mathbf{u}\right) &=  \lVert \mathbf{F} \left(\boldsymbol{\rho}\right) - \mathbf{d} \rVert_2^2 + \lambda \lVert \boldsymbol{\mu} \rVert_1  + \mathbf{u}^\textrm{H} \left[  \boldsymbol{\Theta}\left( \boldsymbol{\rho} \right) - \boldsymbol{\mu} \right] \nonumber \\
&+  \frac{\beta}{2} \lVert  \boldsymbol{\Theta}\left( \boldsymbol{\rho} \right) - \boldsymbol{\mu} \rVert_2^2,
\label{lagrangian}
\end{align}
where $\mathbf{u}$ represents the Lagrangian multiplier or the dual variable, $^\mathrm{H}$ denotes the conjugate transpose operation, and  $\beta > 0$ denotes the augmented Lagrangian parameter.

ADMM consists of the following three updates at the $m^{\textrm{th}}$ iteration:
\begin{align}
\boldsymbol{\mu}^{m+1} &= \argmin_{\boldsymbol{\mu}}  \hspace{0.1cm} \lambda \lVert \boldsymbol{\mu} \rVert_1 + \frac{\beta}{2} \lVert  \boldsymbol{\Theta}\left( \boldsymbol{\rho}^m \right) - \boldsymbol{\mu} + \boldsymbol{\eta}^m\rVert_2^2  \label{update_mu_scaled} \\
\!\!\!\! \boldsymbol{\rho}^{m+1} &= \argmin_{\boldsymbol{\rho}} \hspace{0.1cm} \lVert \mathbf{F} \left(\boldsymbol{\rho}\right) - \mathbf{d} \rVert_2^2 +  \frac{\beta}{2} \lVert  \boldsymbol{\Theta}\left( \boldsymbol{\rho} \right) - \boldsymbol{\mu}^{m+1} + \boldsymbol{\eta}^{m} \rVert_2^2 \label{update_rho_scaled}\\
\boldsymbol{\eta}^{m+1} &= \boldsymbol{\eta}^m + \boldsymbol{\Theta}\left( \boldsymbol{\rho}^{m+1} \right) - \boldsymbol{\mu}^{m+1} \label{update_eta},
\end{align}
where the scaled dual variable $\boldsymbol{\eta}$ is defined by $\boldsymbol{\eta} = \displaystyle \frac{1}{\beta} \mathbf{u}$.

\subsection{Soft thresholding}
The closed-form solution of the optimization problem in (\ref{update_mu_scaled}) is given by
\begin{equation}
\boldsymbol{\mu}^{m+1} = S_{\frac{\lambda}{\beta}} \left( \boldsymbol{\Theta}\left(\boldsymbol{\rho}^m \right) + \boldsymbol{\eta}^{m} \right),
\label{update_mu_analytical}
\end{equation}
where the element-wise soft thresholding operation $S_{\frac{\lambda}{\beta}} (\cdot)$ is  defined as
\begin{equation}
S_{\frac{\lambda}{\beta}} (\alpha) = 
\begin{cases}
\alpha - \frac{\lambda}{\beta}, & \alpha >  \frac{\lambda_1}{\beta}  \\
0, & |\alpha| \leq \frac{\lambda}{\beta} \\
\alpha + \frac{\lambda}{\beta}, & \alpha <  -\frac{\lambda_1}{\beta}
\end{cases}.
\end{equation}

\subsection{Regularized least squares problem}
The update of $\boldsymbol{\rho}$ in (\ref{update_rho_scaled}) takes the form of regularized least squares. The  necessary and sufficient optimality condition is given by
\begin{equation}
\mathbf{A}  \boldsymbol{\rho}^{m+1} = \mathbf{b},
\label{update_rho_optimality}
\end{equation}
where 
\begin{align}
\mathbf{A}  &= \mathbf{F} ^{\mathrm{H}} \mathbf{F}  + \frac{\beta}{2} \boldsymbol{\Theta} ^{\mathrm{H}} \boldsymbol{\Theta}  \label{update_rho_linear_op}, \\
 \mathbf{b} &= \mathbf{F} ^{\mathrm{H}} \mathbf{d} + \frac{\beta}{2} \boldsymbol{\Theta} ^{\mathrm{H}} \left( \boldsymbol{\mu}^{m+1} - \boldsymbol{\eta}^{m}  \right) \label{update_rho_rhs}.
\end{align}
We update $\boldsymbol{\rho}^{m+1}$ by solving the linear equation in (\ref{update_rho_optimality}) using the CG method.

\section{Parallel implementation on TPUs}
In this section, we provide the details on the parallel implementation of ADMM on TPUs,
including the data decomposition, the operations required by the reconstruction and their parallelization,
and the communication strategy of exchanging information among TPU cores.

\begin{figure}[ht]
	\centering
	\includegraphics[width=9cm, height=6.5cm]{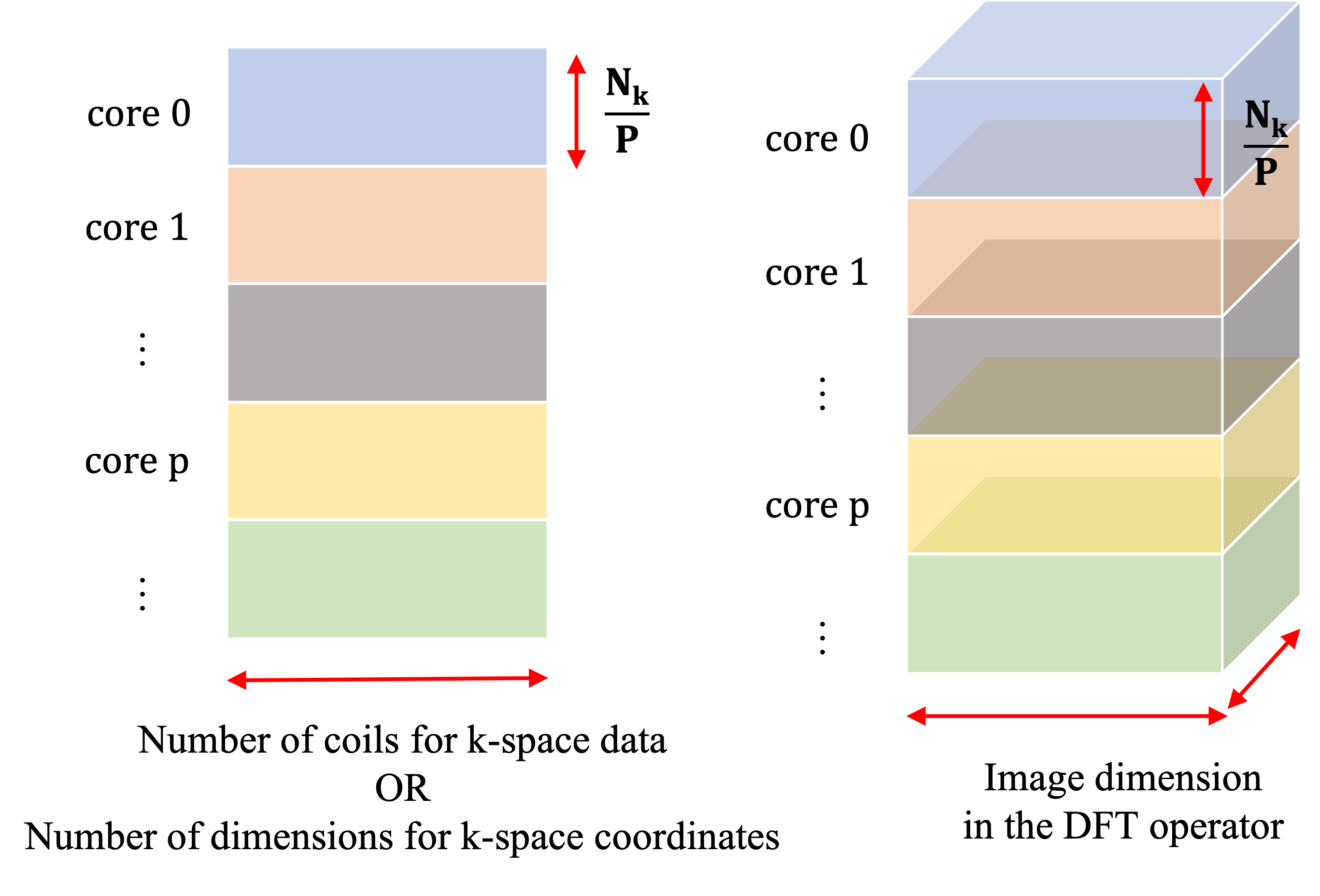}
	\caption{A number of $N_k$ measurements obtained from each of the $N_c$ coils are assigned to $P$ cores, each core contains the measurement data of
		dimension $\frac{N_k}{P} \times N_c$ and  the $k$-space coordinates of dimension $\frac{N_k}{P} \times 2$ for 2D imaging.
		The partition on the $k$-space coordinates enables the generation of the DFT operators on individual cores.
		The partitioned data and operators on different cores are highlighted in different colors.}
	\label{data_decomposition}
\end{figure}

\subsection{Data decomposition}
Data decomposition is applied to the $k$-space such that the measurement data are distributed among TPU cores as shown in Fig.~\ref{data_decomposition}.
If a number of $N_k$ measurements obtained from each of the $N_c$ coils are assigned to $P$ cores, each core contains the measurement data of
dimension $\frac{N_k}{P} \times N_c$.
Similarly, each core also contains partial information of the non-uniform $k$-space coordinates as shown in Fig.~\ref{data_decomposition}.
In the reconstruction of a 2D image, the dimension of the $k$-space coordinates on each core is $\frac{N_k}{P} \times 2$.
There is no data decomposition applied to the image space.

\subsection{DFT and sparsifying transform operators}
The data decomposition applied to the $k$-space enables the generation of the DFT operators in parallel on individual cores, which is shown in Fig.~\ref{data_decomposition}.
The generation of the DFT operator is described in Algorithm~\ref{gen_dft_operator}, which is formulated as tensor contractions with $\texttt{tf.einsum}$.
The sparsifying transform operator is implemented as convolutions with $\texttt{tf.nn.conv1d}$, the performance of which is optimized on TPUs.
Both the DFT and the sparsifying transform operations are performed completely in parallel among the TPU cores.

\begin{algorithm}
	\caption{The generation of the DFT operator on TPUs}
	\label{gen_dft_operator}
	\begin{algorithmic}[1]
		\Function{$\textrm{map\_to\_unit\_circle}$}{$\textrm{k\_r\_product}$}
		\State  \textbf{return}  $\texttt{exp} (-\textrm{i}~2\pi~\textrm{k\_r\_product})$
		\EndFunction
	\end{algorithmic}
	\begin{algorithmic}[1]
		\Function{$\textrm{gen\_dft\_operator}$}{$\textrm{k\_coord}$, $\textrm{image\_coord}$}
		\State $ \textrm{kr\_dim0} \gets \texttt{map\_to\_unit\_circle}(\newline \hspace*{2.3cm}  \texttt{einsum} (\textrm{image\_coord}[0],  ~\textrm{k\_coord}[:, 0]))$
		\State $ \textrm{kr\_dim1} \gets \texttt{map\_to\_unit\_circle}(\newline \hspace*{2.3cm} \texttt{einsum} (\textrm{image\_coord}[1], ~\textrm{k\_coord}[:, 1])) $
		\State $ \textrm{dft\_op} \gets \texttt{einsum} (\textrm{kr\_dim0}, ~\textrm{kr\_dim1}) $
		\State  \textbf{return}  $\textrm{dft\_op}$
		\EndFunction
	\end{algorithmic}
\end{algorithm}

\subsection{One iteration of ADMM and the communication strategy}
One iteration in ADMM consists of three updates in the order of the auxiliary variable, the primal variable or the image intensities, and the dual variable.
The update of the auxiliary variable depends on the sparsifying transform and the soft thresholding operations and has the analytical form  as  in~(\ref{update_mu_analytical}).
The update of the dual variable requires the sparsifying transform as in~(\ref{update_eta}).
Because no partition is applied to the image space, both the sparsifying transform and the soft thresholding operations are performed within individual TPU cores.
Therefore, the update of the auxiliary and the dual variables are completely in parallel among TPU cores.

The image estimate is updated through solving the linear system in (\ref{update_rho_optimality}) with the CG method,
which is described in Algorithm~\ref{cg}.
After one CG iteration, each core only contains the partial image, the exchange of which across cores requires communication.
The communication is to sum the partial image across TPU cores such that all the cores can start a new CG iteration with the same image.
The communication is implemented with $\texttt{tf.cross\_replica\_sum}$, which occurs at the following two places in Algorithm~\ref{cg}:
\begin{itemize}
	\item The communication is required in generating the initial values of the image, or $x0$ in Algorithm~\ref{cg}.
	Note that only one such communication is required prior to the iterative solution process.
	\item The communication with $\texttt{tf.cross\_replica\_sum}$ is embedded in the linear operator $\texttt{linear\_op}$ in Algorithm~\ref{cg}.
	One of the two stopping criteria of the CG solution process is in terms of the squared norm of the residual vector as shown in Algorithm~\ref{cg}.
	The residual vector is obtained by applying the linear operator $\texttt{linear\_op}$ to the image intensities.
	Such a communication is required at every CG iteration.
\end{itemize}

\begin{algorithm}
	\caption{The conjugate gradient method}
	\label{cg}
	\begin{algorithmic}[1]
		\Function{$\textrm{cg\_step}$}{$\texttt{linear\_op}$, $\textrm{r}$, $\textrm{d}$, $\textrm{x}$, $\tau$}
		\State $\textrm{a\_d} \gets \texttt{linear\_op}(\textrm{d})$
		\State $\alpha \gets \texttt{divide}(\tau, ~\texttt{dot\_product}(\textrm{d}, ~\textrm{a\_d}))$
		\State $\textrm{x\_next} \gets \textrm{x} + \alpha ~ \textrm{d} $
		\State $\textrm{r\_next} \gets \textrm{r} - \alpha ~ \textrm{a\_d} $
		\State $\tau\_\textrm{next} \gets \texttt{dot\_product}(\textrm{r\_next}, ~\textrm{r\_next})$
		\State $\beta \gets \texttt{divide}(\tau\_\textrm{next}, ~\tau)$
		\State $\textrm{d\_next} \gets \textrm{r\_next} + \beta~ \textrm{d} $
		\State  \textbf{return}  $\textrm{r\_next}, ~\textrm{d\_next}, ~\textrm{x\_next}, ~\tau\_\textrm{next}$
		\EndFunction
	\end{algorithmic}
	\begin{algorithmic}[1]
		\Function{$\textrm{conjugate\_gradient}$}{$\texttt{linear\_op}$, $\textrm{b}$, $\textrm{x0}$, \newline \hspace*{5cm}  $\textrm{max\_iterations}$,  $\textrm{atol}$} \\
		\Comment{$\textrm{x0}$ contains initial values of $\textrm{x}$ and $\textrm{atol}$ is the absolute tolerance in terms of the norm of the residual vector.}
		\State $\textrm{x} \gets \textrm{x0}$
		\State $\textrm{r} \gets \texttt{linear\_op}(\textrm{x})$
		\State $\textrm{d} \gets \textrm{r}$
		\State $\tau \gets \texttt{dot\_product}( \textrm{r},  \textrm{r})$
		\State $\textrm{i} \gets 0$
		\While {$ \textrm{i}  <  \textrm{max\_iterations} \And \tau > \texttt{square}(\textrm{atol})$}
		\State $\textrm{r\_next}, \textrm{d\_next}, \textrm{x\_next}, \rho\_\textrm{next} \gets \texttt{cg\_step} ( \newline 
		\hspace*{1cm} \texttt{linear\_op}, \textrm{r}, \textrm{d}, \textrm{x}, \tau)$
		\State $\textrm{i} \gets \textrm{i}+ 1$
		\EndWhile
		\State  \textbf{return}  $\textrm{x}$
		\EndFunction
	\end{algorithmic}
\end{algorithm}

As indicated by (\ref{update_rho_linear_op}), the linear operator $\texttt{linear\_op}$ in Algorithm~\ref{cg} consists of two parts:
one corresponding to the DFT operations and the other associated with the sparsifying transform operations.
The communication embedded in $\texttt{linear\_op}$ results from the one corresponding to the DFT operations as shown in Algorithm~\ref{linear_op_cg}.
It is worth mentioning that the encoding with the sensitivity profiles is implemented as tensor contractions with $\texttt{tf.einsum}$
and integrated into the DFT operations.

\begin{algorithm}
	\caption{The linear operator associated with the forward and backward DFT in the conjugate gradient solver on TPUs}
	\label{linear_op_cg}
	\begin{algorithmic}[1]
		\Function{$\textrm{linear\_dft\_op}$}{$\textrm{image}$, $\textrm{sense\_profile}$}
		\State $\textrm{image\_core} \gets \texttt{inv\_DFT}(\texttt{DFT}(\textrm{image}, ~\textrm{sense\_profile}), \newline \hspace*{4.3cm} \textrm{sense\_profile})$
		\State $\textrm{image} \gets \texttt{cross\_replica\_sum}(\textrm{image\_core}) $
		\State  \textbf{return}  $\textrm{image}$
		\EndFunction
	\end{algorithmic}
\end{algorithm}

In addition to the linear operator and the initial values of the image, a vector representing the right-hand side of the linear system shown in (\ref{update_rho_rhs})
is required by the CG solution process at every iteration.
The generation of the right-hand-side vector is performed completely in parallel as both the DFT operator and the $k$-space data are partitioned with respect to
the number of measurement in the $k$-space. The right-hand-side vector on each core can be
generated prior to the CG solution process, which remains the same throughout the CG iterations.

\subsection{Iterations in ADMM}
\begin{algorithm}
	\caption{ADMM on TPUs}
	\label{admm}
	\begin{algorithmic}[1]
		\Function{$\textrm{admm\_step}$}{$\eta$, $\rho$, $\rho_0$}
		\State $\mu\_\textrm{next} \gets \texttt{update\_auxiliary\_var}(\rho, ~\eta)$
		\State $\rho\_\textrm{next} \gets \texttt{update\_primal\_var}(\mu\_\textrm{next}, ~\rho_0, ~\eta)$
		\State $\alpha \gets \texttt{get\_relative\_diff}(\rho\_\textrm{next}, ~\rho)$
		\State $\eta\_\textrm{next} \gets \texttt{update\_dual\_var}(\mu\_\textrm{next}, ~\rho, ~\eta)$
		\State  \textbf{return}  $\eta\_\textrm{next}, ~\rho\_\textrm{next}, ~\alpha$
		\EndFunction
	\end{algorithmic}
	\begin{algorithmic}[1]
		\Function{$\textrm{admm\_reconstruct}$}{$\rho_0$, $\textrm{max\_iterations}$,  $\textrm{rtol}$} \\
		\Comment{$\rho_0$ contains initial values of $\rho$ and $\textrm{rtol}$ is the relative tolerance in terms of the squared norm of the residual.}
		\State $\eta, ~\rho \gets \texttt{get\_initial\_value}(\rho_0)$
		\State $\textrm{i} \gets 0$
		\State $\alpha \gets 1.0$
		\While {$ \textrm{i}  <  \textrm{max\_iterations} \And \alpha > \texttt{square}(\textrm{rtol})$}
		\State $\eta, ~\rho, ~\alpha \gets \texttt{admm\_step} (\eta, ~\rho, ~\rho_0)$
		\State $\textrm{i} \gets \textrm{i}+ 1$
		\EndWhile
		\State  \textbf{return}  $\rho$
		\EndFunction
	\end{algorithmic}
\end{algorithm}
Extending from one iteration to many in ADMM is straightforward and only requires checking the stopping criteria at every iteration as indicated in Algorithm~\ref{admm}.
The stopping criteria is based on the relative difference of the image intensities at two consecutive iterations.
As the image space is not partitioned, checking the stopping criteria is performed completely in parallel among TPU cores.

It can be seen that three major operations are involved in the image reconstruction with ADMM on TPUs, namely,
$\texttt{tf.einsum}$, $\texttt{tf.nn.conv1d}$, and $\texttt{tf.cross\_replica\_sum}$. 
The tensor contractions associated with both the DFT and its inverse operations, as well as the encoding of the sensitivity profiles are implemented with $\texttt{tf.einsum}$.
The sparsifying transform and its adjoint operations based on the finite-difference scheme are implemented with $\texttt{tf.nn.conv1d}$.
The communication among TPU cores is implemented with $\texttt{tf.cross\_replica\_sum}$ and is required at every iteration in the CG solution process to sum the partial image across the cores.
The proposed data decomposition and the parallel implementation of ADMM not only best utilizes TPU's strength in matrix multiplications but also requires minimal communication time,
which leads to very high parallel efficiency.

\section{Accuracy Analysis}
We performed accuracy analysis of the TPU-based reconstructions by comparing with images reconstructed using CPU. Phantom data were acquired using a 3T Siemens MR scanner with a 12-channel phased-array coil. The fully sampled $k$-space data of each channel consisted of 101 radial readouts, each with 128 samples. The images were reconstructed on a $128\times64$ uniform grid. The reconstruction on CPUs was performed with float64, whereas the precision on TPUs was equivalent to float32.

\begin{figure}[ht]
	\centering
	\begin{subfigure}[b]{0.45\linewidth}
		\centering
		\includegraphics[width=3cm, height=6cm]{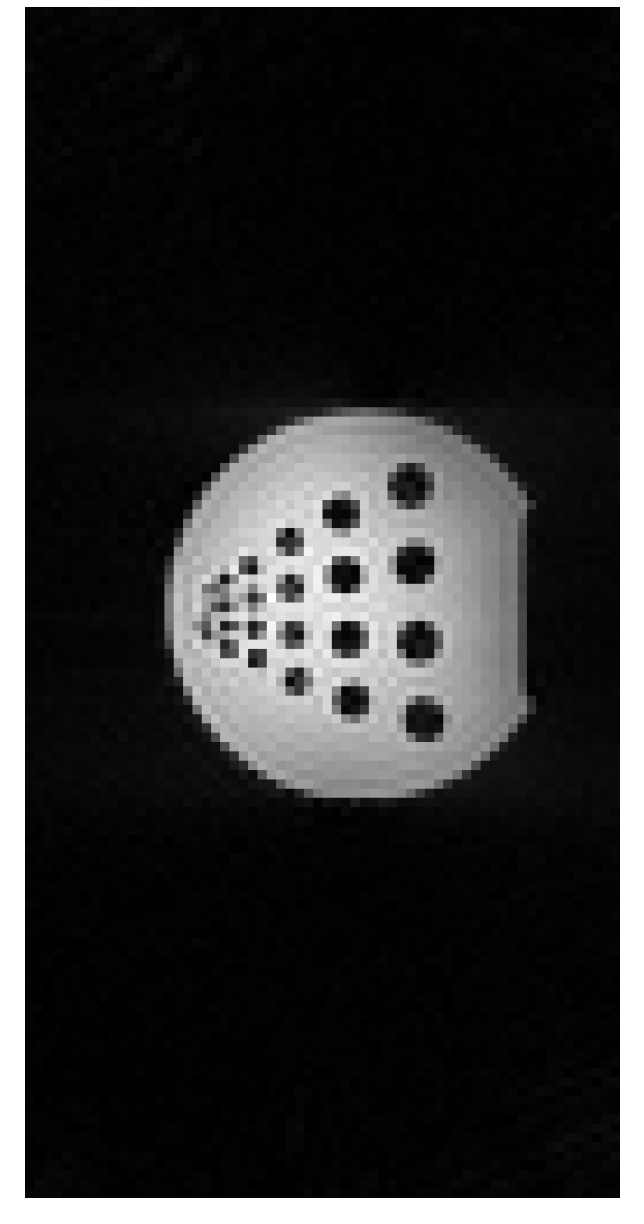}
		\caption{} 
	\end{subfigure}\hfill
	\begin{subfigure}[b]{0.45\linewidth}
		\centering
		\includegraphics[width=3cm, height=6cm]{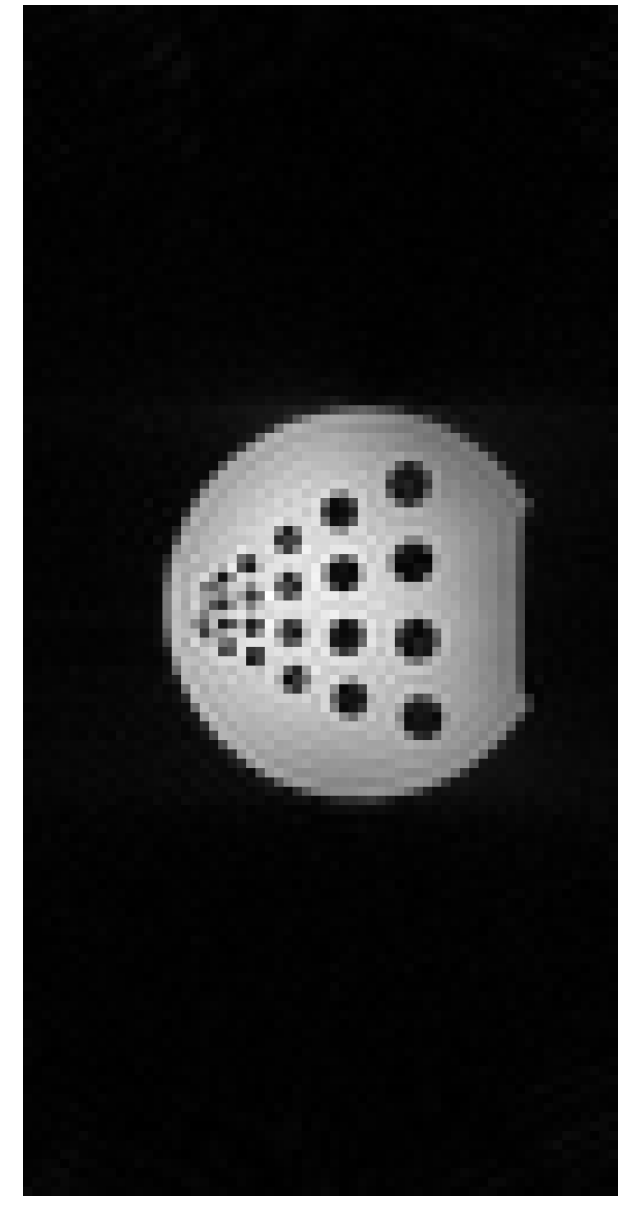}
		\caption{} 
	\end{subfigure}\hfill
	\begin{subfigure}[b]{\linewidth}
		\centering
		\includegraphics[width=8cm, height=6cm]{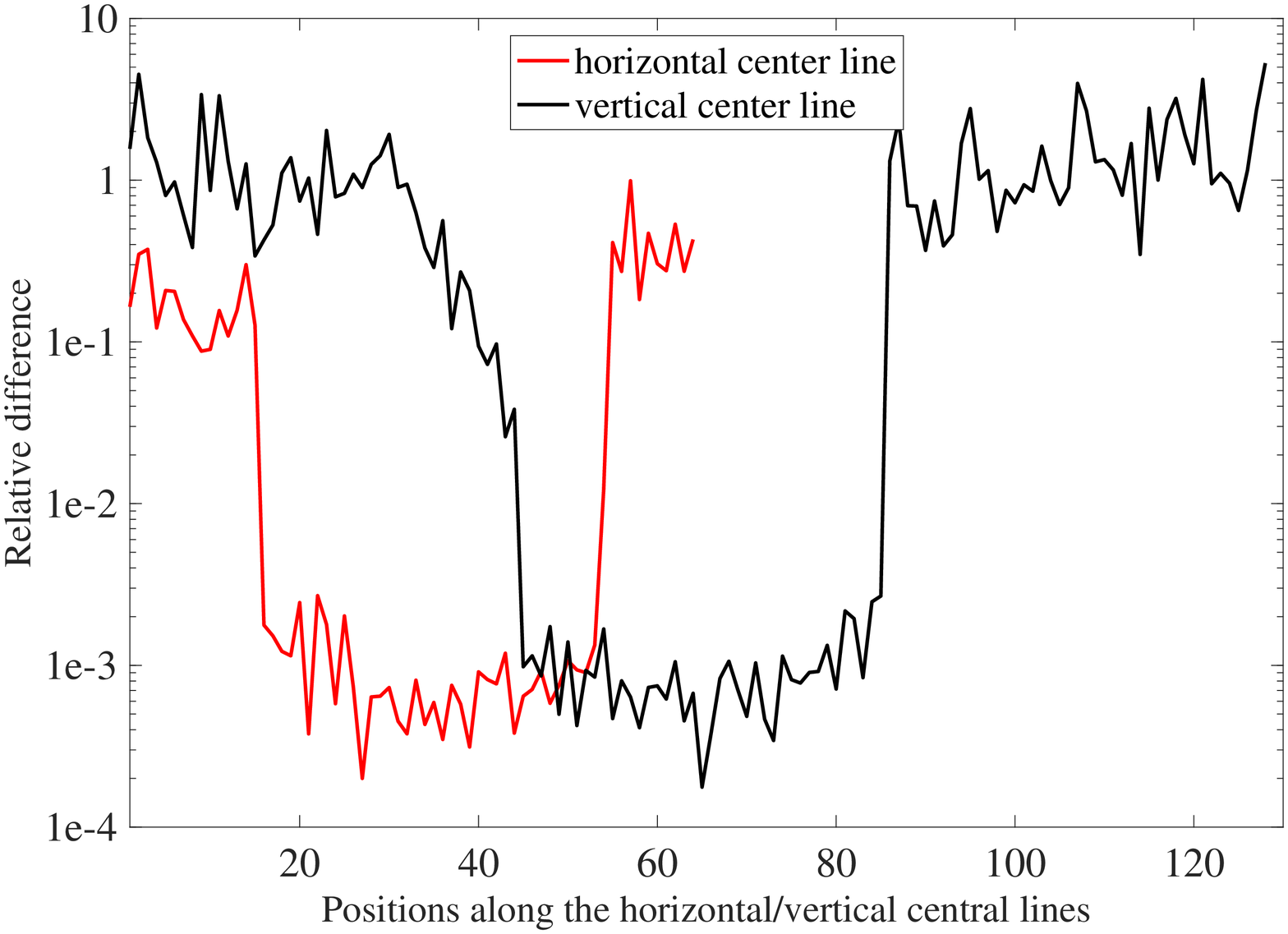}
		\caption{}
	\end{subfigure}
	\caption{Phantom images reconstructed by inverse DFT on (a) CPU and (b) TPUs and (c) plots of the relative difference between the two images along the horizontal and vertical center lines of the image.
	\label{accuracy_dft_cpu_tpu}}
\end{figure}

First, we analyzed the accuracy of the DFT operation on TPUs. Because the DFT operation is repeatedly employed in the iterative image reconstruction using ADMM, it is critical to understand its accuracy. We applied the inverse DFT operation directly to the fully sampled $k$-space data for image reconstruction and compared the results obtained by TPUs with those by CPU. Figures \ref{accuracy_dft_cpu_tpu}(a) and (b) show the reconstructed images by CPU and TPUs, respectively. Figure~\ref{accuracy_dft_cpu_tpu}(c) plots the relative difference of the reconstructed images along a horizontal (red) and vertical (black) line both at the center of the image, respectively. As can be seen, the relative difference is about $0.1\%$ for the voxels within the phantom. Large relative difference occurs on the noise background as expected.

\begin{table*}[]
	\renewcommand{\arraystretch}{1.5}
	\caption{Parameters used by ADMM in the reconstruction of the phantom image shown in Fig.~\ref{accuracy_admm_cpu_tpu}. }
	\centering
	\begin{tabular}{|c|c|c|c|c|c|}
		\hline
		\multicolumn{4}{|c|}{ADMM}  & \multicolumn{2}{c|}{CG}   \\ \hline
		\begin{tabular}[c]{@{}c@{}}Regularization\\ parameter $\lambda$\end{tabular} & \begin{tabular}[c]{@{}c@{}}Augmented Lagrangian\\ parameter $\beta$\end{tabular} & \begin{tabular}[c]{@{}c@{}} Relative \\ tolerance \end{tabular} & \begin{tabular}[c]{@{}c@{}}Maximum number of\\ iterations\end{tabular} & \begin{tabular}[c]{@{}c@{}} Absolute \\ tolerance \end{tabular}  & \begin{tabular}[c]{@{}c@{}}Maximum number of\\ iterations\end{tabular} \\ \hline
		$10^{-7}$  & 1.0 & $10^{-4}$  & 5  & $10^{-6}$ & 20  \\ \hline
	\end{tabular}
	\label{parameters_accuracy_benchmark}
\end{table*}

\begin{figure}[ht]
	\centering
	\begin{subfigure}[b]{0.33\linewidth}
		\centering
		\includegraphics[width=3cm, height=6cm]{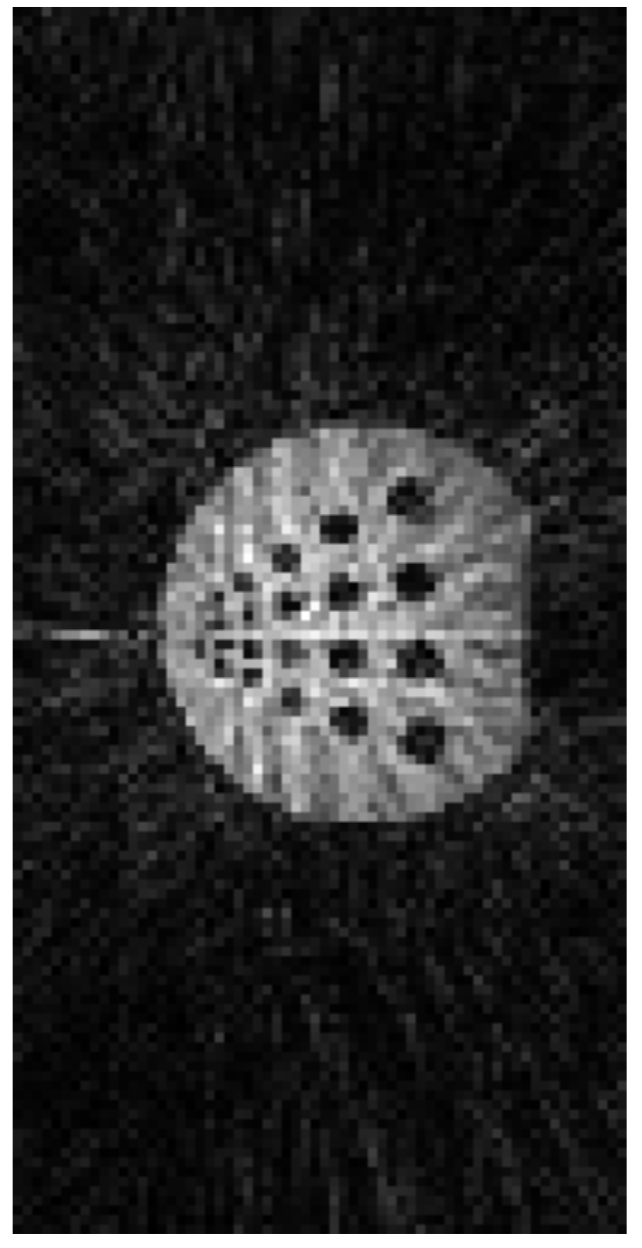}
		\caption{} 
	\end{subfigure}\hfill
	\begin{subfigure}[b]{0.33\linewidth}
		\centering
		\includegraphics[width=3cm, height=6cm]{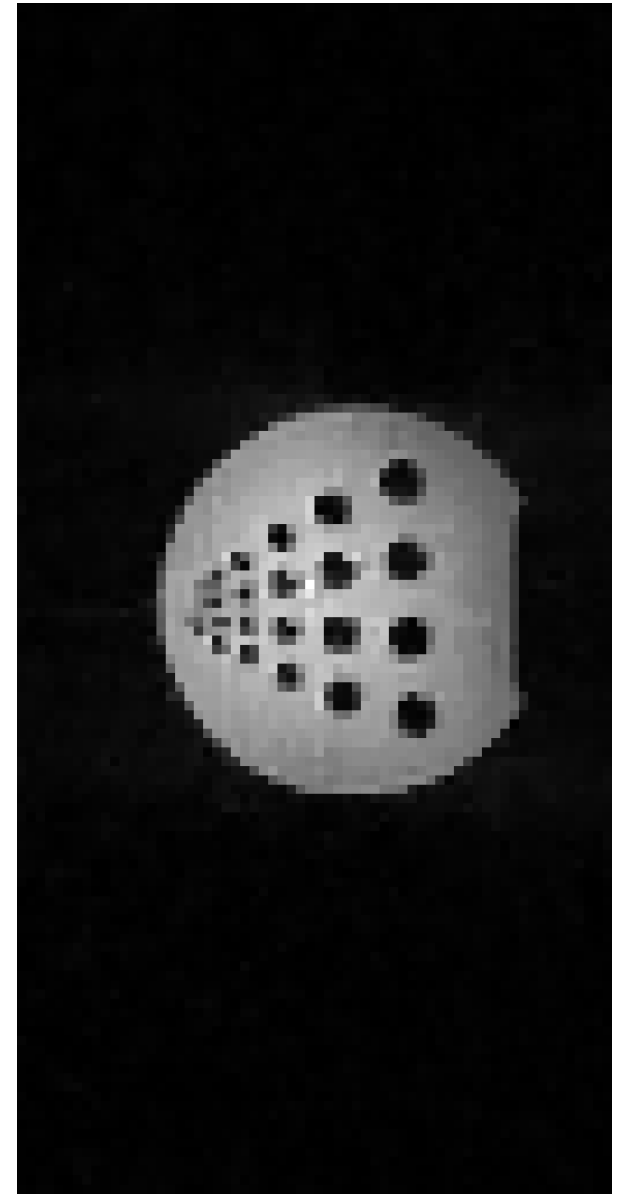}
		\caption{} 
	\end{subfigure}\hfill
	\begin{subfigure}[b]{0.33\linewidth}
		\centering
		\includegraphics[width=3cm, height=6cm]{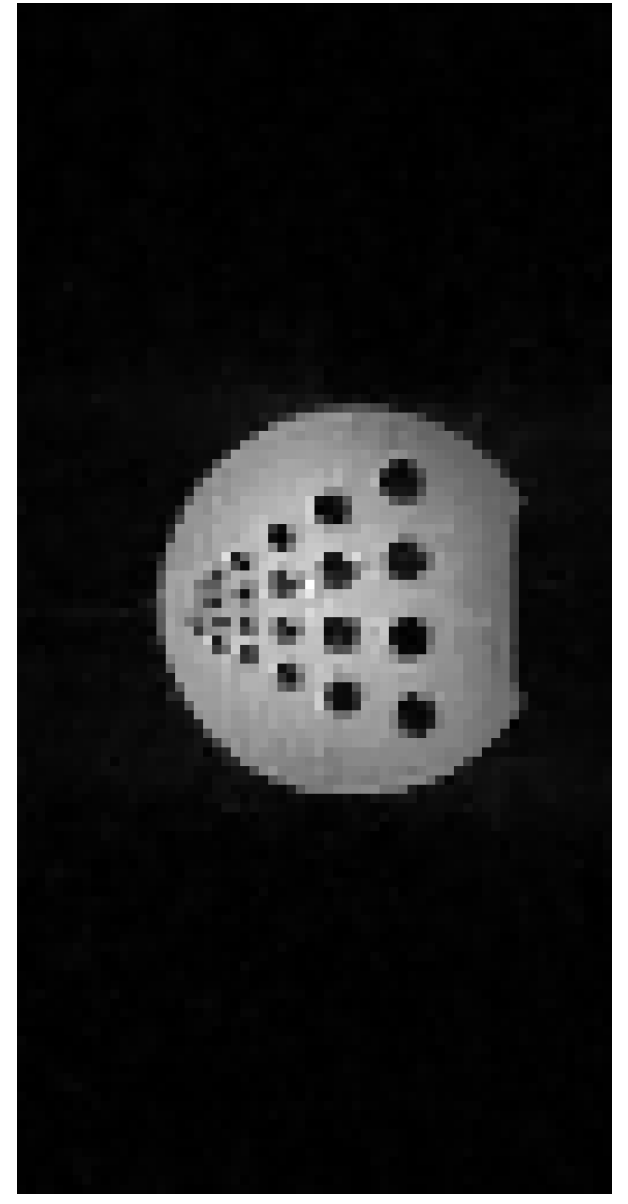}
		\caption{} 
	\end{subfigure}\hfill
	\begin{subfigure}[b]{\linewidth}
		\centering
		\includegraphics[width=8cm, height=6cm]{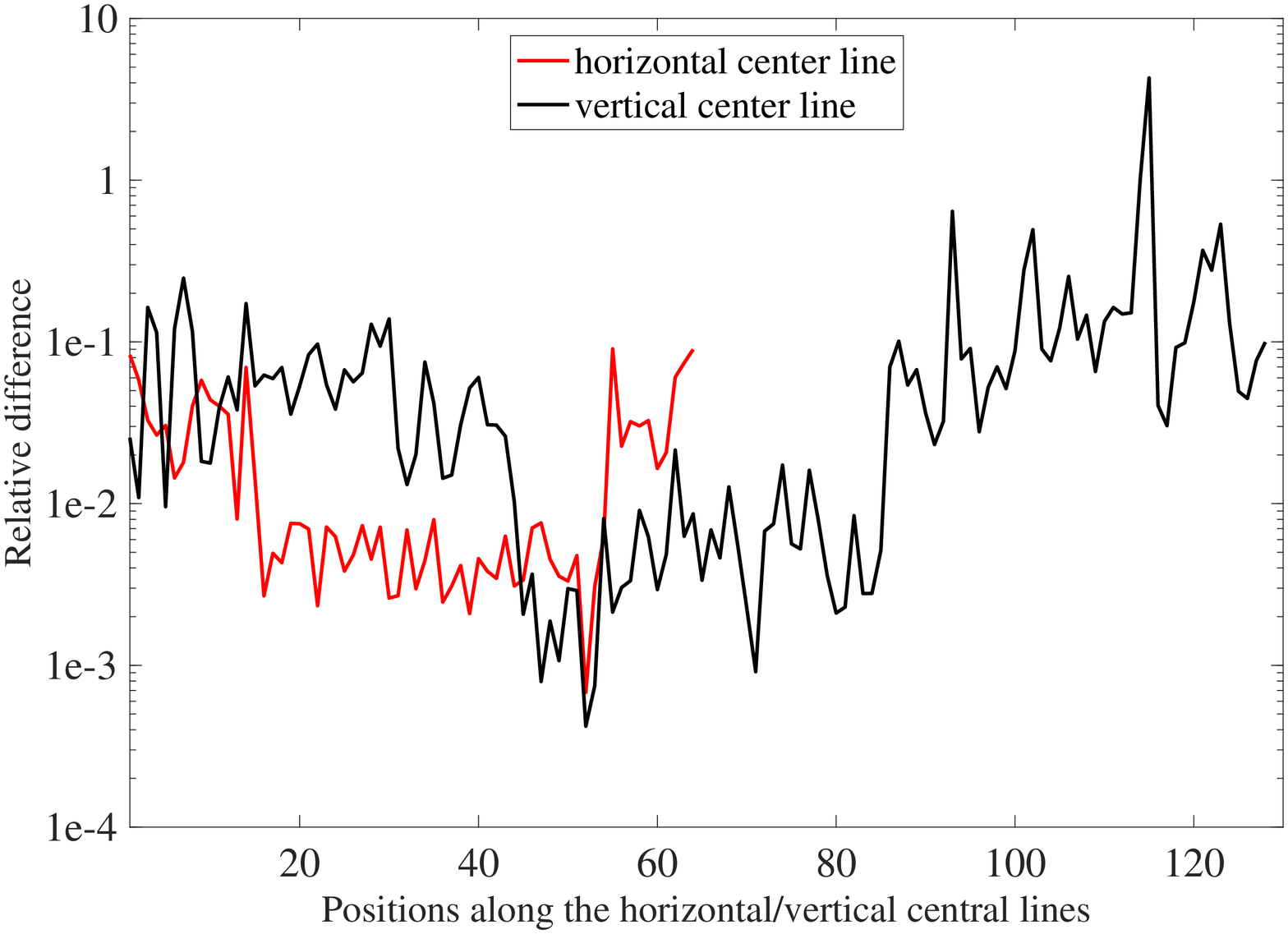}
		\caption{}
	\end{subfigure}
	\caption{Reconstructed phantom images by using retrospectively undersampled data (a) with a single inverse DFT operation and (b) by the ADMM algorithm on CPU and (c) TPUs; and (d) the relative difference between the two images in (b) and (c) along the horizontal and vertical center lines of the image.}
	\label{accuracy_admm_cpu_tpu}
\end{figure}

Second, we analyzed the accuracy of the iterative image reconstruction on TPUs. We performed the ADMM-based image reconstructions, i.e., (\ref{update_mu_scaled}) to (\ref{update_eta}), on CPU and TPUs, respectively. The $k$-space data were retrospectively underampled with a undersampling factor of eight to demonstrate the capability of compressed sensing in accelerating MR. The total number of $k$-space measurements was 19,968 with 1,664 samples or 13 radial readouts for each coil and 12 coils in total. The images were reconstructed on a $1024\times512$ uniform grid. The hyper-parameters of the ADMM algorithms are listed in Table~\ref{parameters_accuracy_benchmark}.
Figure~\ref{accuracy_admm_cpu_tpu}(a) shows the phantom image reconstructed by inverse DFT, displaying expected aliasing artifacts due to sparse sampling in the $k$-space.
Figures \ref{accuracy_admm_cpu_tpu}(b) and (c) show the reconstructed images on CPU and TPUs, respectively. Figure~\ref{accuracy_admm_cpu_tpu}(d) plots the relative difference between the reconstructed images in Figs.\ \ref{accuracy_admm_cpu_tpu}(b) and (c) along a horizontal (red) and vertical (black) line at the center of the image, respectively. As can be seen, the relative difference is about $1\%$ for the voxels within the phantom. In the ADMM-based image reconstruction, there were over 100 calls for both the DFT and its inverse operations. The accumulated numerical errors over the iterations explain the larger difference between TPUs- and CPU-based image reconstruction than the case of direct inverse DFT reconstruction. Note that the image reconstruction error for compressed sensing MR is often around $5\%$. Therefore, the $1\%$ additional error due to float32-based computations on TPUs is acceptable for most MR applications.

\section{Parallel Efficiency Analysis}
In this section, we demonstrate the high parallel efficiency of the image reconstruction on TPUs through numerical examples. The strong scaling analysis was adopted to understand the parallel efficiency, in which the problem size remained the same and the number of TPU cores used for the reconstruction varied. The computation time was measured with the TPU profiling tool \cite{tpuv3}, which also provides information on the hardware utilization and the computation time associated with individual operations.

\begin{figure}[t!]
	\includegraphics[width=8cm, height=6cm]{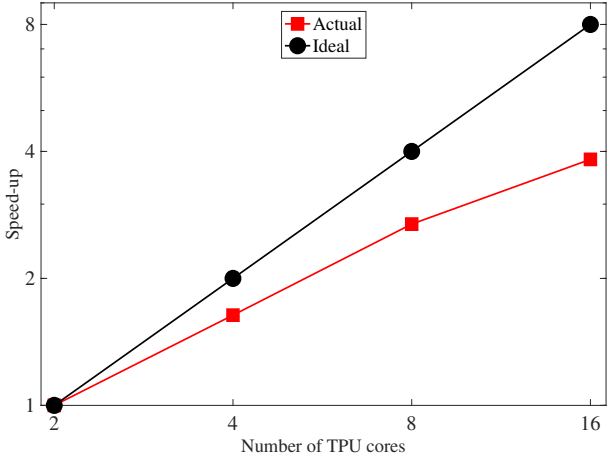}
	\centering
	\caption{The speed-up of reconstructing an image of size $128 \times 64$ with up to 16 TPU cores.
		The number of $k$-space measurements is 19,968 with 1,664 samples for each coil and 12 coils in total.}
	\label{speedup_image_size_128_64}
\end{figure}

\begin{figure}[t!]
	\includegraphics[width=8cm, height=6cm]{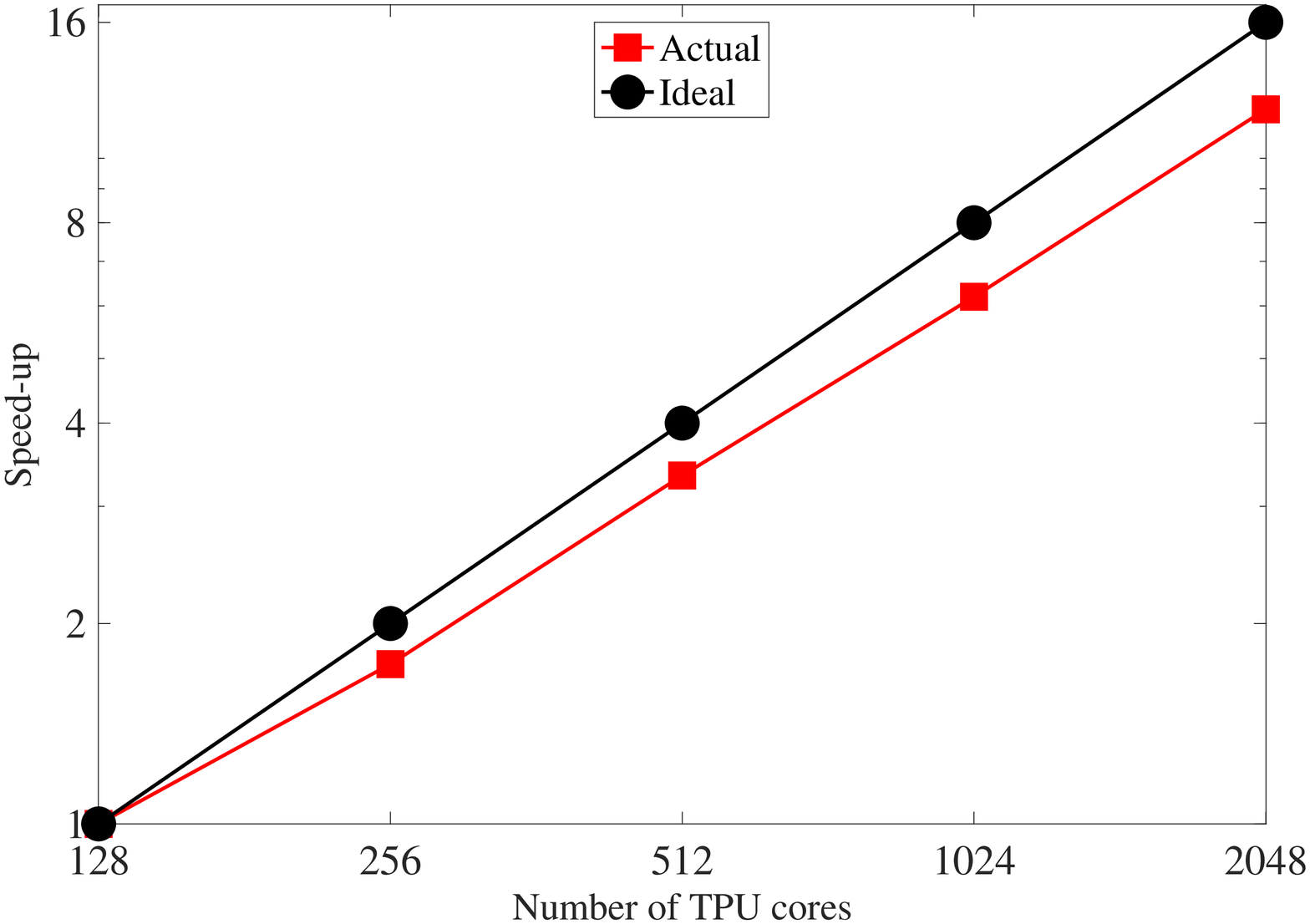}
	\centering
	\caption{The speed-up of reconstructing an image of size $1024 \times 512$ with up to 2048 TPU cores.
		The number of $k$-space measurements is 1,241,076 with 103,424 samples for each coil and 12 coils in total.}
	\label{speedup_image_size_1024_512}
\end{figure}

The first numerical example for the parallel efficiency analysis is the one used for understanding the reconstruction accuracy as shown in Fig.~\ref{accuracy_admm_cpu_tpu}. With the problem size remaining the same, the number of TPU cores used for computation increased from two to 16. Figure\ \ref{speedup_image_size_128_64} plots the corresponding speed-up defined as:
\begin{equation}
\textrm{speed-up} = \frac{T_{2}}{T_{N_{\textrm{core}}}},
\label{speed_up_2core}
\end{equation}
where $T_2$ denotes the computation time with two TPU cores and $T_{N_{\textrm{core}}}$ represents the computation time with $N_{\textrm{core}}$ cores. As a reference, the ideal speed-up corresponding to the linear scaling is also depicted in Fig.~\ref{speedup_image_size_128_64}. Because the problem size is relatively small, when the number of TPU cores increases from four to eight, the gain of the speedup saturates, which can be seen in Fig.~\ref{speedup_image_size_128_64}. 
The image reconstruction time was 136.3 ms with two TPU cores and 35.6 ms with 16 cores.

The second numerical example for the parallel efficiency analysis is the reconstruction of a much larger image. Phantom data were acquired by using the same scanner as in Section V but with 804 radial readouts, each with 1024 samples. The fully sampled $k$-space data were then retrospectively undersampled by a factor of eight, resulting in a total number of 1,241,076 $k$-space measurements (103,424 measurements per coil, 12 coils in total). The parameters used by ADMM are the same as those in Table~\ref{parameters_accuracy_benchmark}. Figure~\ref{speedup_image_size_128_64} plots the corresponding speed-up defined as:
\begin{equation}
\textrm{speed-up} = \frac{T_{128}}{T_{N_{\textrm{core}}}},
\label{speed_up_128core}
\end{equation}
where $T_{128}$ denotes the computation time with 128 TPU cores and $T_{N_{\textrm{core}}}$ represents the computation time with $N_{\textrm{core}}$ cores. The ideal speed-up of linear scaling is also depicted in Fig.~\ref{speedup_image_size_1024_512}. It can be seen that the achieved speed-up is very close to that of the linear scaling. As a reference, the computation time was 3.387 seconds with 128 TPU cores and 0.286 seconds with 2048 cores. Both numerical examples demonstrate very high parallel efficiency of the image reconstruction with ADMM on TPUs. Given the parallel efficiency demonstrated through examples in Figs.~\ref{speedup_image_size_128_64} and \ref{speedup_image_size_1024_512}, a very large problem can be solved with similar amount of computation time to that of a much smaller problem if more TPU cores are used.

\section{Discussion}
The image reconstruction on TPUs was implemented with precision float32. Compared to the CPU-based image reconstruction with precision float64, our accuracy analysis shows that using float32 instead of float64 led to about $0.1\%$ difference for non-iterative DFT-based image reconstruction and $1\%$ difference for the ADMM-based iterative image reconstruction. Considering the $5\%$ reconstruction error in compressed sensing MR, using float32 is acceptable for most MR applications.

\begin{table*}[]
	\renewcommand{\arraystretch}{1.5}
	\caption{Computation time of the ADMM-based reconstruction on CPU, GPU, and TPUs.
	The non-uniform Fourier transform is implemented as DFT on TPUs and NUFFT on CPU and GPU. The CPU used for the comparison is Intel(R) Xeon(R) Silver 4110 8-core 2.10 GHz and the GPU is Nvidia Tesla V100 SXM2. One TPU unit or board consists of four chips, which is shown in Fig.\ \ref{tpu_chip_board_pod}(b).}
	\centering
	\begin{tabular}{|c|c|c|c|c|c|}
	\hline
	\multicolumn{2}{|c|}{}                                                                          & \multicolumn{4}{c|}{Computation time (seconds)}                 \\ \hline
	\multicolumn{2}{|c|}{Hardware}                                                                  & CPU    & GPU   & \multicolumn{2}{c|}{TPU (number of TPU units)} \\ \hline
	\multicolumn{2}{|c|}{\begin{tabular}[c]{@{}c@{}}Non-uniform\\ Fourier\\ Transform\end{tabular}} & NUFFT  & NUFFT & \multicolumn{2}{c|}{DFT}                       \\ \hline
	\multirow{2}{*}{\begin{tabular}[c]{@{}c@{}}Image \\ Size\end{tabular}}       & 128 X 64         & 2.38   & 1.17  & 0.14 (1/4 unit)       & 0.036 (two units)      \\ \cline{2-6} 
	& 1024 X 512       & 139.88 & 2.24  & 3.39 (16 units)       & 0.29 (128 units)       \\ \hline
	\end{tabular}
	\label{computation_time_comparision}
\end{table*}

In this work, the non-uniform Fourier transform and its adjoint  were implemented as matrix multiplications, which has computational complexity of $\mathcal{O}(N^2)$ and $N$ is the number of pixels in an image. There is a computationally more efficient algorithm to perform the Fourier transform known as the non-uniform fast Fourier transform (NUFFT) \cite{Fessler2003}, which is a fast Fourier transform (FFT)-based algorithm and has the computational complexity of $\mathcal{O}(N\log N)$. 
There are existing CPU and GPU-based packages for the reconstruction of non-Cartesian $k$-space data that are implemented with NUFFT.  
As shown in Table \ref{computation_time_comparision}, we performed a preliminary comparison in terms of the runtimes for the image reconstructions on CPU, GPU, and TPUs.
Except that the DFT operator in the ADMM algorithm was replaced with the NUFFT operator on CPU and GPU,
the rest of the reconstruction method among CPU, GPU, and TPUs remains the same. 
The image reconstruction on CPU and GPU was implemented with SigPy \cite{Ong2019}.
The CPU used for the comparison is Intel(R) Xeon(R) Silver 4110 8-core 2.10 GHz and the GPU is Nvidia Tesla V100 SXM2.
The runtimes for reconstructing the $128\times64$ image from the undersampled data (the task in Fig.~\ref{speedup_image_size_128_64}) are 2.38 seconds on CPU, 1.17 seconds on GPU, 0.14 seconds with $\frac{1}{4}$ TPU unit, and 0.036 seconds with two units, respectively.
Given the difference in the computational complexity of the Fourier transform, the reconstruction on TPUs achieves satisfying acceleration.
The runtimes for reconstructing the $1024\times512$ image from the undersampled data (the task in Fig.~\ref{speedup_image_size_1024_512}) are 139.88 seconds on CPU, 2.24 seconds on GPU, 3.39 seconds with 16 TPU units, and 0.29 seconds with 256 units, respectively.
For this large image size, the computational complexity reduction offered by NUFFT starts to overcome the TPU acceleration.
Our work on implementing NUFFT on TPUs is on-going and will be reported when it is ready.
Considering the difference of the computational complexity between DFT and NUFFT, we believe the above comparison demonstrates that the reconstruction on TPUs is promising for accelerating MRI process with its high parallel efficiency.

\section{Conclusion}
In this work, we propose using TPU---the domain-specific hardware accelerator originally developed for deep learning applications---to accelerate MR image reconstruction. 
The ADMM-based iterative image reconstruction algorithm was implemented on TPUs using TensorFlow because of its rich set of functionalities for scientific computing and simplicity in expressing parallel algorithms. 
The data decomposition was applied to $k$-space data for the parallel reconstruction on TPUs. Correspondingly, the DFT operator arising from the non-Cartesian sampling of the $k$-space was partitioned and assigned to individual TPU cores. There were three major TensorFlow operations used in the TPU-based image reconstruction, namely, $\texttt{tf.einsum}$, $\texttt{tf.nn.conv1d}$, $\texttt{tf.cross\_replica\_sum}$. The DFT operator and its inverse, as well as the encoding of the sensitivity profiles were all treated as tensor contractions with $\texttt{tf.einsum}$. The sparsifying transform and its adjoint operators based on the finite-difference scheme were handled with $\texttt{tf.nn.conv1d}$. The operations of $\texttt{tf.einsum}$ and $\texttt{tf.nn.conv1d}$ can be performed in a highly efficient manner on TPUs. In addition, all these operations formulated as $\texttt{tf.einsum}$ and $\texttt{tf.nn.conv1d}$ were localized within individual TPU cores and performed completely in parallel. Because of the data decomposition, only a partial image was obtained on each TPU core, the exchange of which requires communication among TPU cores. The communication was implemented with $\texttt{tf.cross\_replica\_sum}$ and was responsible of summing the partial image across TPU cores at each iteration. Given the fact that TPUs are connected directly to each other with dedicated high-speed interconnects of very low latency, the communication was also highly efficient. The proposed parallel computing algorithm not only best utilizes TPU's strength in matrix multiplications but also requires minimal communication time, leading to very high parallel efficiency, which are demonstrated through numerical examples.
In conclusion, we have demonstrated the potential of using TPUs for accelerating MR image reconstruction and the achieved high parallel  efficiency. 

\section*{Acknowledgment}
We would like to thank John Anderson, Tao Wang, Yusef Shafi, James Lottes, Damien Pierce, Qing Wang, and Lily Hu at Google for valuable discussions and helpful comments.

Thibault Marin, Yue Zhuo, and Chao Ma were supported in part by the National Institute of Health under award: T32EB013180, R01CA165221, R01HL118261, R01HL137230 and P41EB022544.

\bibliographystyle{IEEEtran}
\bibliography{bibliography}

\begin{thebibliography}{10}
\providecommand{\url}[1]{#1}
\csname url@samestyle\endcsname
\providecommand{\newblock}{\relax}
\providecommand{\bibinfo}[2]{#2}
\providecommand{\BIBentrySTDinterwordspacing}{\spaceskip=0pt\relax}
\providecommand{\BIBentryALTinterwordstretchfactor}{4}
\providecommand{\BIBentryALTinterwordspacing}{\spaceskip=\fontdimen2\font plus
\BIBentryALTinterwordstretchfactor\fontdimen3\font minus
  \fontdimen4\font\relax}
\providecommand{\BIBforeignlanguage}[2]{{%
\expandafter\ifx\csname l@#1\endcsname\relax
\typeout{** WARNING: IEEEtran.bst: No hyphenation pattern has been}%
\typeout{** loaded for the language `#1'. Using the pattern for}%
\typeout{** the default language instead.}%
\else
\language=\csname l@#1\endcsname
\fi
#2}}
\providecommand{\BIBdecl}{\relax}
\BIBdecl

\bibitem{sodickson1997simultaneous}
D.~K. Sodickson and W.~J. Manning, ``Simultaneous acquisition of spatial
  harmonics {(SMASH)}: fast imaging with radiofrequency coil arrays,''
  \emph{Magnetic resonance in medicine}, vol.~38, no.~4, pp. 591--603, 1997.

\bibitem{pruessmann1999sense}
K.~P. Pruessmann, M.~Weiger, M.~B. Scheidegger, and P.~Boesiger, ``{SENSE}:
  sensitivity encoding for fast {MRI},'' \emph{Magnetic Resonance in Medicine:
  An Official Journal of the International Society for Magnetic Resonance in
  Medicine}, vol.~42, no.~5, pp. 952--962, 1999.

\bibitem{pruessmann2006encoding}
K.~P. Pruessmann, ``Encoding and reconstruction in parallel {MRI},'' \emph{NMR
  in Biomedicine: An International Journal Devoted to the Development and
  Application of Magnetic Resonance In vivo}, vol.~19, no.~3, pp. 288--299,
  2006.

\bibitem{candes2006robust}
E.~J. Cand{\`e}s, J.~Romberg, and T.~Tao, ``Robust uncertainty principles:
  Exact signal reconstruction from highly incomplete frequency information,''
  \emph{IEEE Transactions on information theory}, vol.~52, no.~2, pp. 489--509,
  2006.

\bibitem{donoho2006compressed}
D.~L. Donoho, ``Compressed sensing,'' \emph{IEEE Transactions on information
  theory}, vol.~52, no.~4, pp. 1289--1306, 2006.

\bibitem{candes2006near}
E.~J. Cand{\`e}s and T.~Tao, ``Near-optimal signal recovery from random
  projections: Universal encoding strategies?'' \emph{IEEE transactions on
  information theory}, vol.~52, no.~12, pp. 5406--5425, 2006.

\bibitem{lustig2007sparse}
M.~Lustig, D.~Donoho, and J.~M. Pauly, ``Sparse {MRI}: The application of
  compressed sensing for rapid {MR} imaging,'' \emph{Magnetic Resonance in
  Medicine: An Official Journal of the International Society for Magnetic
  Resonance in Medicine}, vol.~58, no.~6, pp. 1182--1195, 2007.

\bibitem{liang2007ps}
Z.-P. Liang, ``Spatiotemporal imaging with partially separable functions,'' p.
  988–991, 2007.

\bibitem{stone2007gpumr}
S.~S. Stone, H.~Yi, J.~P. Haldar, W.-m.~W. Hwu, B.~P. Sutton, and Z.-p. Liang,
  ``How {GPU}s can improve the quality of magnetic resonance imaging,'' in
  \emph{In The First Workshop on General Purpose Processing on Graphics
  Processing Units}, 2007.

\bibitem{pratx2011gpumedicalphysics}
\BIBentryALTinterwordspacing
G.~Pratx and L.~Xing, ``{GPU} computing in medical physics: a review,''
  \emph{Medical Physics}, vol.~38, no.~5, pp. 2685--2697, 2011. [Online].
  Available: \url{https://doi.org/10.1118/1.3578605}
\BIBentrySTDinterwordspacing

\bibitem{eklund2013medicalgpu}
\BIBentryALTinterwordspacing
A.~Eklund, P.~Dufort, D.~Forsberg, and S.~M. LaConte, ``Medical image
  processing on the {GPU} - past, present and future,'' \emph{Medical Image
  Analysis}, vol.~17, no.~8, pp. 1073 -- 1094, 2013. [Online]. Available:
  \url{https://doi.org/https://doi.org/10.1016/j.media.2013.05.008}
\BIBentrySTDinterwordspacing

\bibitem{despres2017reviewgpurecon}
\BIBentryALTinterwordspacing
P.~Despr{\'e}s and X.~Jia, ``A review of {GPU}-based medical image
  reconstruction,'' \emph{Physica Medica: European Journal of Medical Physics},
  vol.~42, pp. 76--92, Oct 2017. [Online]. Available:
  \url{https://doi.org/10.1016/j.ejmp.2017.07.024}
\BIBentrySTDinterwordspacing

\bibitem{wang2018surveygpumr}
\BIBentryALTinterwordspacing
H.~Wang, H.~Peng, Y.~Chang, and D.~Liang, ``A survey of {GPU}-based
  acceleration techniques in mri reconstructions,'' \emph{Quantitative Imaging
  in Medicine and Surgery}, vol.~8, no.~2, 2018. [Online]. Available:
  \url{http://qims.amegroups.com/article/view/18832}
\BIBentrySTDinterwordspacing

\bibitem{Zhuo2011}
Y.~Zhuo, X.-L. Wu, J.~P. Haldar, T.~Marin, W.-m.~W. Hwu, Z.-P. Liang, and B.~P.
  Sutton, \emph{{U}sing {GPU}s to {A}ccelerate {A}dvanced {MRI}
  {R}econstruction with {F}ield {I}nhomogeneity {C}ompensation}, ser.
  Applications of GPU Computing Series.\hskip 1em plus 0.5em minus 0.4em\relax
  Morgan Kaufmann, 2011, pp. 709--722.

\bibitem{stone2008nufft}
\BIBentryALTinterwordspacing
S.~S. Stone, J.~P. Haldar, S.~C. Tsao, W.-m.~W. Hwu, B.~P. Sutton, and Z.-P.
  Liang, ``Accelerating advanced mri reconstructions on {GPU}s,'' \emph{Journal
  of Parallel and Distributed Computing}, vol.~68, no.~10, pp. 1307 -- 1318,
  2008, general-Purpose Processing using Graphics Processing Units. [Online].
  Available: \url{https://doi.org/https://doi.org/10.1016/j.jpdc.2008.05.013}
\BIBentrySTDinterwordspacing

\bibitem{soerensen2008nufft}
\BIBentryALTinterwordspacing
T.~S. S{\"o}rensen, T.~Schaeffter, K.~{\O}. Noe, and M.~S. Hansen,
  ``Accelerating the nonequispaced fast fourier transform on commodity graphics
  hardware,'' \emph{IEEE Transactions on Medical Imaging}, vol.~27, no.~4, pp.
  538--547, April 2008. [Online]. Available:
  \url{https://doi.org/10.1109/TMI.2007.909834}
\BIBentrySTDinterwordspacing

\bibitem{yang2009nufft}
\BIBentryALTinterwordspacing
Z.~Yang and M.~Jacob, ``Efficient {NUFFT} algorithm for non-cartesian {MRI}
  reconstruction,'' in \emph{IEEE International Symposium on Biomedical
  Imaging}, June 2009, pp. 117--120. [Online]. Available:
  \url{https://doi.org/10.1109/ISBI.2009.5192997}
\BIBentrySTDinterwordspacing

\bibitem{Zhuo2010d}
Y.~Zhuo, X.-L. Wu, J.~P. Haldar, W.-m.~W. Hwu, Z.-P. Liang, and B.~P. Sutton,
  ``{A}ccelerating iterative field-compensated {MR} image reconstruction on
  {GPU}s,'' in \emph{2010 {IEEE} {I}nternational {S}ymposium on {B}iomedical
  {I}maging: {F}rom {N}ano to {M}acro}, 2010, pp. 820--823.

\bibitem{boyd2011distributed}
S.~Boyd, N.~Parikh, E.~Chu, B.~Peleato, and J.~Eckstein, ``Distributed
  optimization and statistical learning via the alternating direction method of
  multipliers,'' \emph{Foundations and Trends{\textregistered} in Machine
  learning}, vol.~3, no.~1, pp. 1--122, 2011.

\bibitem{hansen2008cartktsense}
\BIBentryALTinterwordspacing
M.~S. Hansen, D.~Atkinson, and T.~S. S{\"o}rensen, ``Cartesian sense and k-t
  sense reconstruction using commodity graphics hardware,'' \emph{Magnetic
  Resonance in Medicine}, vol.~59, no.~3, pp. 463--468, 2008. [Online].
  Available: \url{https://doi.org/10.1002/mrm.21523}
\BIBentrySTDinterwordspacing

\bibitem{murphy2012l1spirit}
\BIBentryALTinterwordspacing
M.~Murphy, M.~Alley, J.~Demmel, K.~Keutzer, S.~Vasanawala, and M.~Lustig,
  ``Fast $\ell_1$ -spirit compressed sensing parallel imaging mri: Scalable
  parallel implementation and clinically feasible runtime,'' \emph{IEEE
  Transactions on Medical Imaging}, vol.~31, no.~6, pp. 1250--1262, June 2012.
  [Online]. Available: \url{https://doi.org/10.1109/TMI.2012.2188039}
\BIBentrySTDinterwordspacing

\bibitem{smith2012bregman}
\BIBentryALTinterwordspacing
D.~S. Smith, J.~C. Gore, T.~E. Yankeelov, and E.~B. Welch,
  ``\BIBforeignlanguage{eng}{Real-time compressive sensing mri reconstruction
  using {GPU} computing and split bregman methods},''
  \emph{\BIBforeignlanguage{eng}{International journal of biomedical imaging}},
  vol. 2012, pp. 864\,827--864\,827, 2012, 22481908[pmid]. [Online]. Available:
  \url{https://doi.org/10.1155/2012/864827}
\BIBentrySTDinterwordspacing

\bibitem{nam2013cs3drad}
\BIBentryALTinterwordspacing
S.~Nam, M.~Ak{\c{c}}akaya, T.~Basha, C.~Stehning, W.~J. Manning, V.~Tarokh, and
  R.~Nezafat, ``Compressed sensing reconstruction for whole-heart imaging with
  3d radial trajectories: a graphics processing unit implementation,''
  \emph{Magnetic Resonance in Medicine}, vol.~69, no.~1, pp. 91--102, 2013.
  [Online]. Available: \url{https://doi.org/10.1002/mrm.24234}
\BIBentrySTDinterwordspacing

\bibitem{chang2017cs3d}
\BIBentryALTinterwordspacing
C.-H. Chang, X.~Yu, and J.~X. Ji, ``Compressed sensing {MRI} reconstruction
  from 3d multichannel data using {GPU}s,'' \emph{Magnetic Resonance in
  Medicine}, vol.~78, no.~6, pp. 2265--2274, 2017. [Online]. Available:
  \url{https://doi.org/10.1002/mrm.26636}
\BIBentrySTDinterwordspacing

\bibitem{schaetz2012multigpulib}
S.~Schaetz and M.~Uecker, ``A multi-{GPU} programming library for real-time
  applications,'' in \emph{Algorithms and Architectures for Parallel
  Processing}, Y.~Xiang, I.~Stojmenovic, B.~O. Apduhan, G.~Wang, K.~Nakano, and
  A.~Zomaya, Eds.\hskip 1em plus 0.5em minus 0.4em\relax Berlin, Heidelberg:
  Springer Berlin Heidelberg, 2012, pp. 114--128.

\bibitem{freiberger2013agile}
\BIBentryALTinterwordspacing
M.~Freiberger, F.~Knoll, K.~Bredies, H.~Scharfetter, and R.~Stollberger, ``The
  agile library for biomedical image reconstruction using {GPU} acceleration,''
  \emph{Computing in Science Engineering}, vol.~15, no.~1, pp. 34--44, Jan
  2013. [Online]. Available: \url{https://doi.org/10.1109/MCSE.2012.40}
\BIBentrySTDinterwordspacing

\bibitem{wu2011impatient}
\BIBentryALTinterwordspacing
X.-L. Wu, J.~Gai, F.~Lam, M.~Fu, J.~P. Haldar, Y.~Zhuo, Z.-P. Liang, W.-M.~W.
  Hwu, and B.~P. Sutton, ``{I}mpatient {MRI}: {I}llinois {M}assively {P}arallel
  {A}cceleration {T}oolkit for image reconstruction with enhanced throughput in
  {MRI},'' in \emph{{IEEE} {I}nternational {S}ymposium on {B}iomedical
  {I}maging}.\hskip 1em plus 0.5em minus 0.4em\relax IEEE, 2011, pp. 69--72,
  exported from refbase (https://dell-desktop:81/refbase/show.php?record=2870),
  last updated on Tue, 07 Apr 2020 14:32:34 -0400. [Online]. Available:
  \url{https://doi.org/10.1109/ISBI.2011.5872356}
\BIBentrySTDinterwordspacing

\bibitem{gai2013moreimpatient}
\BIBentryALTinterwordspacing
J.~Gai, N.~Obeid, J.~L. Holtrop, X.-L. Wu, F.~Lam, M.~Fu, J.~P. Haldar,
  W.-m.~W. Hwu, Z.-P. Liang, and B.~P. Sutton, ``More impatient: a
  gridding-accelerated toeplitz-based strategy for non-cartesian
  high-resolution 3d mri on {GPU}s,'' \emph{Journal of Parallel and Distributed
  Computing}, vol.~73, no.~5, pp. 686 -- 697, 2013. [Online]. Available:
  \url{https://doi.org/https://doi.org/10.1016/j.jpdc.2013.01.001}
\BIBentrySTDinterwordspacing

\bibitem{cerjanic2016powergrid}
A.~Cerjanic, J.~L. Holtrop, G.~C. Ngo, B.~Leback, G.~Arnold, M.~Van~Moer,
  G.~LaBelle, J.~A. Fessler, and B.~P. Sutton, ``Powergrid: A open source
  library for accelerated iterative magnetic resonance image reconstruction,''
  in \emph{Proc. Intl. Soc. Mag. Res. Med}, vol. 525, 2016.

\bibitem{uecker2019}
\BIBentryALTinterwordspacing
M.~Uecker and J.~Tamir, ``mrirecon/bart: version 0.5.00,'' Aug. 2019. [Online].
  Available: \url{https://doi.org/10.5281/zenodo.3376744}
\BIBentrySTDinterwordspacing

\bibitem{stoica2017berkeley}
I.~Stoica, D.~Song, R.~A. Popa, D.~Patterson, M.~W. Mahoney, R.~Katz, A.~D.
  Joseph, M.~Jordan, J.~M. Hellerstein, J.~E. Gonzalez \emph{et~al.}, ``A
  {B}erkeley view of systems challenges for {AI},'' \emph{arXiv preprint
  arXiv:1712.05855}, 2017.

\bibitem{jouppi2017datacenter}
N.~P. Jouppi, C.~Young, N.~Patil, D.~Patterson, G.~Agrawal, R.~Bajwa, S.~Bates,
  S.~Bhatia, N.~Boden, A.~Borchers \emph{et~al.}, ``In-datacenter performance
  analysis of a tensor processing unit,'' in \emph{2017 ACM/IEEE 44th Annual
  International Symposium on Computer Architecture (ISCA)}.\hskip 1em plus
  0.5em minus 0.4em\relax IEEE, 2017, pp. 1--12.

\bibitem{tpuv3}
\BIBentryALTinterwordspacing
Cloud {TPU}s. [Online]. Available: \url{https://cloud.google.com/tpu/}
\BIBentrySTDinterwordspacing

\bibitem{yang2019high}
\BIBentryALTinterwordspacing
K.~Yang, Y.-F. Chen, G.~Roumpos, C.~Colby, and J.~Anderson, ``High performance
  {M}onte {C}arlo simulation of {I}sing model on {TPU} clusters,'' in
  \emph{Proceedings of the International Conference for High Performance
  Computing, Networking, Storage and Analysis}, ser. SC '19.\hskip 1em plus
  0.5em minus 0.4em\relax ACM, 2019, pp. 83:1--83:15. [Online]. Available:
  \url{http://doi.acm.org/10.1145/3295500.3356149}
\BIBentrySTDinterwordspacing

\bibitem{belletti2019tensor}
F.~Belletti, D.~King, K.~Yang, R.~Nelet, Y.~Shafi, Y.-F. Chen, and J.~Anderson,
  ``Tensor processing units for financial {M}onte {C}arlo,'' \emph{arXiv
  preprint arXiv:1906.02818}, 2019.

\bibitem{lu2020large}
T.~Lu, Y.-F. Chen, B.~Hechtman, T.~Wang, and J.~Anderson, ``Large-scale
  discrete {F}ourier transform on {TPU}s,'' \emph{arXiv preprint
  arXiv:2002.03260}, 2020.

\bibitem{jouppi2017quantifying}
\BIBentryALTinterwordspacing
N.~Jouppi. (2017) Quantifying the performance of the {TPU}, our first machine
  learning chip. [Online]. Available:
  \url{https://cloud.google.com/blog/products/gcp/quantifying-the-performance-of-the-tpu-our-first-machine-learning-chip}
\BIBentrySTDinterwordspacing

\bibitem{wu2016google}
Y.~Wu, M.~Schuster, Z.~Chen, Q.~V. Le, M.~Norouzi, W.~Macherey, M.~Krikun,
  Y.~Cao, Q.~Gao, K.~Macherey \emph{et~al.}, ``Google's neural machine
  translation system: Bridging the gap between human and machine translation,''
  \emph{arXiv preprint arXiv:1609.08144}, 2016.

\bibitem{bfloat16}
\BIBentryALTinterwordspacing
Using bfloat16 with {T}ensor{F}low models. [Online]. Available:
  \url{https://cloud.google.com/tpu/docs/bfloat16}
\BIBentrySTDinterwordspacing

\bibitem{ramani2010parallel}
S.~Ramani and J.~A. Fessler, ``Parallel {MR} image reconstruction using
  augmented lagrangian methods,'' \emph{IEEE Transactions on Medical Imaging},
  vol.~30, no.~3, pp. 694--706, 2010.

\bibitem{Fessler2003}
J.~A. Fessler and B.~P. Sutton, ``{N}onuniform fast {F}ourier transforms using
  min-max interpolation,'' \emph{IEEE Transactions on Signal Processing},
  vol.~51, no.~2, pp. 560--574, 2003.

\bibitem{Ong2019}
F.~Ong and M.~Lustig, ``{S}ig{P}y: {A} {P}ython {P}ackage for {H}igh
  {P}erformance {I}terative {R}econstruction,'' in \emph{{P}roc. {ISMRM}},
  2019.

\end{thebibliography}

\end{document}